\documentclass[letter]{aa} 

\usepackage{natbib}
\bibpunct{(}{)}{;}{a}{}{,} 
\usepackage{graphicx}
\usepackage{txfonts}
\newcommand{\fat}[1]{{#1}}

\begin{document}

   \title{The mass of Albireo Aa and the nature of Albireo AB
   }

   \subtitle{New aspects from Gaia DR2}

   \author{U. Bastian
          \inst{1}
          \and
          R. Anton\inst{2}
          }

   \institute{Zentrum f\"ur Astronomie (Center for Astronomy), Heidelberg University,
              M\"onchhofstr. 14, D-69120 Heidelberg\\
              \email{bastian@ari.uni-heidelberg.de}
         \and           
Internationale Amateursternwarte e.V. (International Amateur Observatory Ass.), Grevenkamp 5, D-24161 Altenholz \\
             \email{rainer.anton@ki.comcity.de}
             }

   \date{Received October xx, 2018; accepted November yy, 2018}

 
  \abstract
   {}
   {We aim to clarify the nature of Albireo AB and specifically to decipher whether it is an optical or physical pair. We also try to determine the mass of Albireo Aa.}
   {We scrutinize and compare the available absolute astrometric data (from Hipparcos and Gaia DR2) of Albireo A and B, and  we investigate the relative orbit of the pair Albireo \fat{Aa,Ac} using orbit solutions based on ground-based interferometric measurements.}
   {The mass of Albireo Aa (K3\,II) is surprisingly small; only an upper limit of about 0.7\,$M_\odot$ could be derived. The systemic proper motion of \fat{Aa,Ac} differs from that of component B by about 10\,mas/year with an uncertainty of less than 2\,mas/year. Albireo AB is therefore most probably an optical double.}
   {Specific astrometric and spectroscopic follow-up observations clarifying the surprising mass estimate for Albireo Aa are recommended.}


   \maketitle
%

\section{Introduction}

A precise knowledge of the brightest binary stars is one major observational basis for
critical testing of our stellar models, and thus to gain a profound understanding of stellar physics. This is especially non-trivial when red
giants are involved, as the more advanced stages of stellar evolution present many complications, such as understanding internal
mixing processes and their consequences 
\citep[see] [and references therein] {Schroder1997} 
or the evolution of angular momentum and stellar activity, to name only two
interesting aspects. While for simple geometric reasons eclipsing
binaries are the most unambigious source of precise information on stellar
physical parameters including mass, nearby giants with well-determined
astrometric orbits can serve the same purpose. In this respect, the
well-known double star Albireo is not only a beautiful object for
public observations and astronomical outreach work, but is also an important astrophysical research target. Albireo AB, $\beta$\,Cyg, is a wide pair of about 35\arcsec\,separation. The brighter component A is a close binary with a separation of order 0.4\arcsec\,between components Aa and Ac. 

As a funny side remark we mention that the present short study of the system was prompted by a posting on Aug 13, 2018, on the online news and social networking service “Twitter” which raised considerable interest in the astronomy-interested public. It reads {\it “… one of the most famous and beautiful double stars in the sky has been exposed as a fake (optical) double. ESA’s Gaia reveals the two stars are 60 light years apart, just coincidentally lined up}”.

Looking up the relevant parallax data from Gaia Data Release 2 (DR2), this statement turns out to be based on doubtful information. The published values and formal errors of the DR2 parallaxes for Albireo A and B, namely $\varpi_A=9.95\pm 0.60$mas\footnote{mas = milliarcsec} and $\varpi_B=8.38\pm 0.17$mas, lead to a parallax difference of $\Delta\varpi=1.57$mas with a formal uncertainty of $\sigma_{\Delta\varpi}=0.62$mas. This means that the difference is only marginally significant at $2.5\sigma$. But the actual significance must be considered even lower than this formal one, for two reasons: firstly, Gaia can measure extremely bright Albireo A ($G$ magnitude 2.43, and V=3.085) only in the form of strongly saturated images 
\citep[see] [and references therein] {Gaia2016}. For Gaia DR2 
\citep{Gaia2018,Lindegren2018} these could not yet be sufficiently calibrated\footnote{Note that Gaia was designed and announced for a bright magnitude limit of $G=5.7$, i.e.~Albireo A is over 20 times brighter than this limit.} astrometrically. This can be seen from the overall error distribution of stars with $G<5$ in DR2 as well as from some strange individual results for such very bright stars. Secondly, Albireo A itself is a tight binary, almost resolved at the angular resolution of Gaia, and with very significant orbital acceleration over the two years of mission entering Gaia DR2. The non-pointlike nature of the object confers additional complications in the pre-reduction of the individual astrometric measurements, and the orbital acceleration may directly impair the two-year parallax adjustment (Albireo, like all other 1.7 billion stars in DR2, was astrometrically solved assuming a constant proper motion).   
Unfortunately, all these specific problems cannot be securely quantified, but they give good reason to suspect that the actual uncertainty on the parallax is somewhat larger than the formal one. A closer look at this prominent triple-star system in the light of Gaia DR2 therefore appears worthwhile.


\section{The orbit of the pair Albireo \fat{Aa,Ac}}

\begin{figure}
   \centering
      \includegraphics[width=6.6cm]{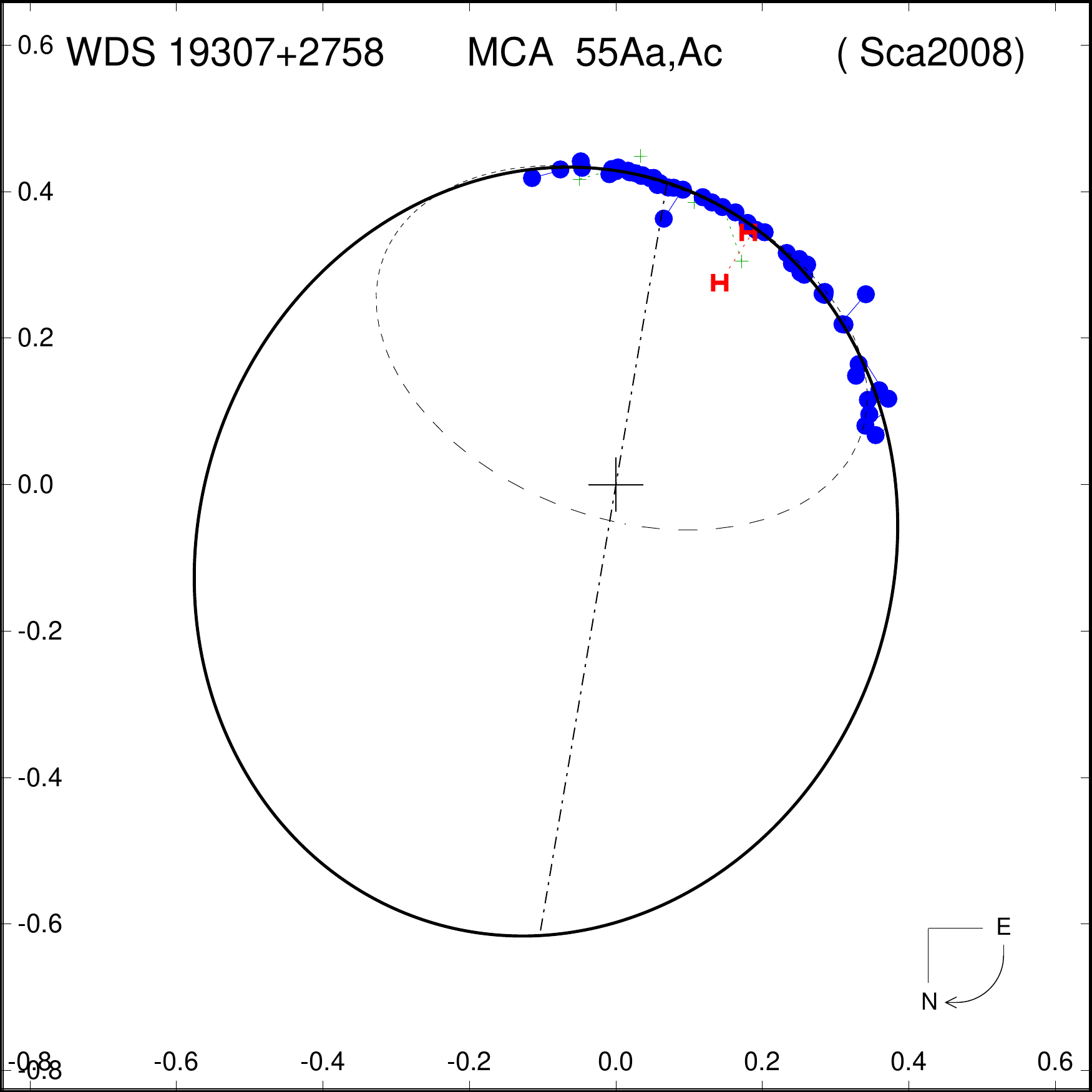}
   \caption{Relative orbit of Albireo Ac (with respect to Albireo Aa) according to \citet{Scardia2008}, including line of nodes \fat{(dash-dotted line), focal-point location (cross) and all available observations (including the few most recent ones that were not used for the orbit solution). The plot is in equatorial (J2000) coordinates; the scale is in arcseconds. Blue dots are interferometric, green crosses are visual, and red H marks Hipparcos measurements. \newline The \citet{Roberts2018} orbit is indicated for comparison (dashed ellipse). The slightly increasing residuals at the end of the observations, that is at the end of the covered orbital arc, motivated the orbit solution of \citet{Roberts2018}.} }
              \label{Scardia-orbit}%
    \end{figure}

\begin{figure}
   \centering
   \includegraphics[width=6.6cm]{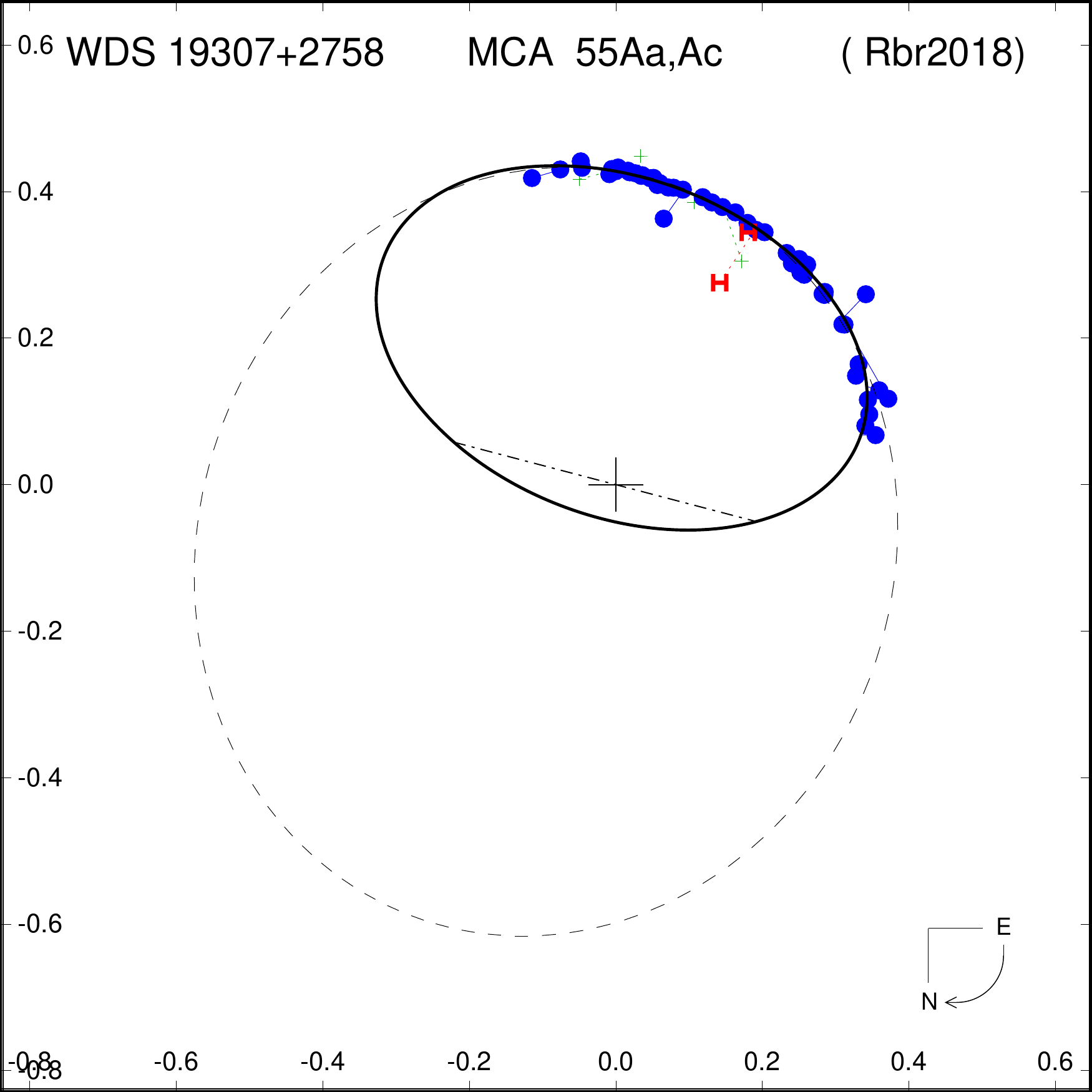}
   \caption{Relative orbit of Albireo Ac (wrt Albireo Aa) according to \citet{Roberts2018}, including line of nodes (dash-dotted line), focal-point location (cross) and observations used for the orbit solution. \fat{Symbols and orientation are as in Fig.\,\ref{Scardia-orbit}. This plot differs from the original figure of 
\citet{USNO18} solely by the addition of the \citet{Scardia2008}~orbit for comparison (dashed ellipse).} }
              \label{Roberts-orbit}%
    \end{figure}

Albireo A consists of a bright giant (Albireo Aa, spectral type K3\,II) and a main-sequence star (Albireo Ac, B9\,V).\footnote{\fat{There are vague indications of another component, dubbed Albireo Ab. It has been detected only twice \citep{Bonneau1980, Prieur2002}, and in both cases only marginally. We ignore it in the present paper because this history gives significant reason to doubt its existence.}} 
The pair has been resolved interferometrically at various epochs, and there are two published orbit solutions which differ significantly even though they are based on mostly the same set of measurements. These two solutions by 
\citet{Scardia2008} (period 214 years) and 
\citet{Roberts2018} (period 69 years) are shown in Figs.\,\ref{Scardia-orbit} and \,\ref{Roberts-orbit}, respectively. As we shall see, the former fits well to the Hipparcos and Gaia data --- although with an astrophysical surprise --- while the latter is completely excluded by the measured Gaia DR2 position and proper motion. \citet{Roberts2018} mention that their orbit solution leads to a mass sum of 87~$M_\odot$  when the parallax from the Hipparcos re-reduction \citep{Leeuwen2007} is applied. This is unreasonable for a K3\,II\,+\,B9\,V pair. The \citet{Scardia2008}~orbit, on the contrary, leads to much more plausible mass sums of 5.7~$M_\odot$ and 3.3~$M_\odot$ if the Hipparcos and the Gaia DR2 parallaxes are applied, respectively.

\begin{figure}
   \centering
   \includegraphics[width=8cm]{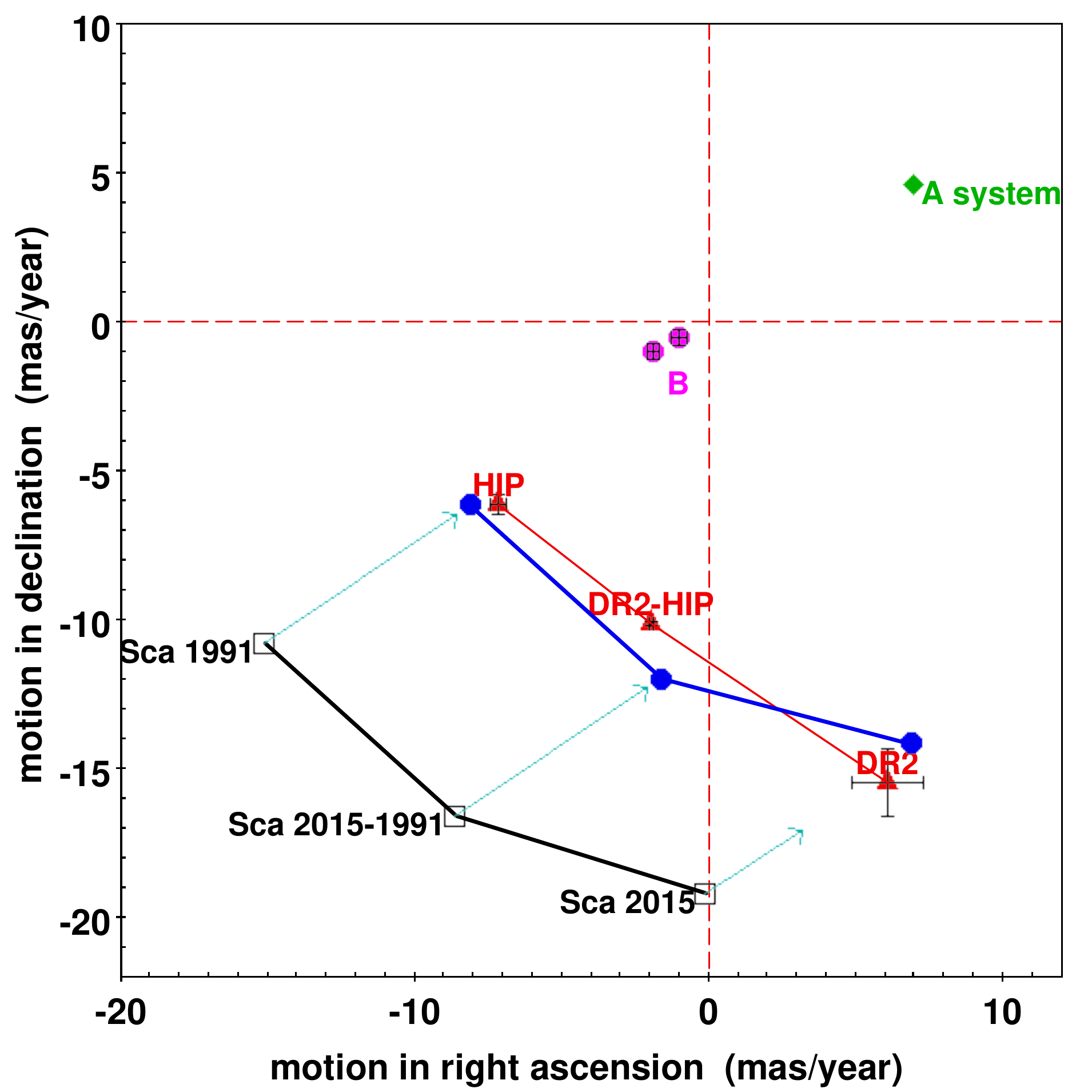}
   \caption{Combined vector point diagram of all proper motions, both relative and absolute, relevant for the present study. Horizontal axis is $\mu_\alpha\cos\delta$ and vertical axis is $\mu_\delta$ in ICRS orientation. The labelled red triangles denote the absolute measurements of the proper motion of Albireo A by Hipparcos (epoch J1991.25), by Gaia DR2 (epoch J2015.5), and by the difference between the Gaia DR2 and Hipparcos positions. The open black squares denote the relative motion of Aa as derived from the \citet{Scardia2008}~orbit solution for the pair \fat{Aa,Ac} (i.e.~from a mirror image of Fig.~\ref{Scardia-orbit}) for the same epochs. The blue circles show the same data, but shifted by an assumed systemic motion of \fat{Aa,Ac} of (+7.0,+4.6)\,mas/year to give a best-possible fit to the absolute motions of A (red triangles).
That absolute systemic motion is indicated by the green diamond, to be compared with the two violet circles showing the absolute proper motions of B measured by Hipparcos (lower left) and Gaia DR2 (upper right). Their error bars are about the size of the symbols.}    
              \label{central}%
    \end{figure}

There are three measured values for the {\it absolute} proper motion of Albireo A: the quasi-instantaneous J1991.25 motion measured by Hipparcos, the quasi-instantaneous J2015.5 motion measured by Gaia DR2, and the mean motion between J1991.25 and J2015.5 as measured by the position difference between Hipparcos and Gaia DR2. These are plotted as labelled red triangles in Fig.\,\ref{central}, along with error bars representing the formal uncertainties from the relevant star catalogues. The error bars for the mean motion between J1991.25 and J2015.5 are smaller than the symbol, those for J1991.25 are about the size of the symbol. 

The same figure also displays the {\it relative} motion of Albireo Aa with respect to Ac, derived for the same epochs from the \citet{Scardia2008} orbit solution (black open squares at lower left). As can be seen, the pattern is very similar, with an obvious offset due to the systemic velocity of the pair \fat{Aa,Ac}. Within the uncertainties of both the absolute motions and the relative orbit solution, the two patterns are in agreement. This is more clearly shown by the blue filled circles in Fig.~\ref{central} which represent the relative motion shifted by an assumed \fat{Aa,Ac} systemic motion of (+7.0/+4.6)~mas/year. This systemic motion is a best-fit estimate carried out by eye. It is not useful to try a more formal fit, as the uncertainty of the orbit solution cannot be well quantified. Nevertheless, at first sight the blue symbols in Fig.\,\ref{central} can be interpreted as mutually confirming the absolute motion measurements and the  relative orbit solution by \citet{Scardia2008}. 

The alternative orbit solution of \citet{Roberts2018} gives almost the same relative motion for J1991.25 (not surprisingly, as most of the observations used cluster around that epoch), but it predicts a grossly deviating evolution of that motion. The mean motion between J1991.25 and J2015.5 is already outside the right-hand border of Fig.~\ref{central}, and the predicted instantaneous motion at J2015.5 is at the very far right, at about (+95,+20)\,mas/year, with a tremendous acceleration towards the north. This is in stark contrast to the Gaia DR2 data.   
         
\section{The mass of Albireo Aa}

There is an astrophysical surprise in the agreement between the blue circles and the red triangles in Fig.\,\ref{central}: The shape of the red and blue patterns should be identical, but the amplitude of the red pattern should reflect the mass ratio of Albireo Aa (K3\,II) and Ac (B9\,V). More precisely, the amplitude ratio of the absolute and relative motion patterns reflects the ratio of the absolute and relative orbital semi-major axis of the measured component Aa. Denoting the relative semi-major axis as $a$, and the absolute semi-major axis as $a_{Aa}$, these are related to the masses $M_{Aa}$ and $M_{Ac}$ by
\begin{equation}
M_{Aa}=M_{Ac} (a/a_{Aa} - 1)
.\end{equation}  

Taking the agreement between the observed $a$ and $a_{Aa}$ at face value means that the mass of the giant star Aa must be much smaller than that of the main-sequence star Ac. The ratio of the pattern amplitudes in Fig.\,\ref{central} is surely consistent with 1.0 (formally leading to zero $M_{Aa}$) and 0.9, and less so with 0.8. Values less than 0.8 are implausible from Fig.\,\ref{central}. Assuming $M_{Ac}=3\,M_\odot$ for the B~star, and a plausible ratio of~0.9, we find $M_{Aa}=0.3\,M_\odot$. With an assumed lower limit for the ratio of 0.8 we correspondingly find a rough upper limit of  $M_{Aa}=0.75\,M_\odot$.

Such small masses for giant stars do occur in close binaries after mass overflow to the other component. But the orbital separation of Albireo Aa and Ac is of the order of 40~au. Therefore a significant mass transfer seems to be excluded, and the low mass of Albireo Aa is surprising. 

The mass discrepancy is very large. Using the known parallax, spectral type, and V magnitude to compare the giant Aa with evolutionary tracks on the Hertzsprung-Russell diagram, several solar masses are inferred. This is true whatever tracks are used. Taking the tracks of 
\citet{Schroder1997} as an example, the resulting mass is over 5\,$M_\odot$.
We note that any possible contribution from the light of the B star (component Ac) to the measured proper motion of Aa (the giant) would even reduce the upper mass limit derived above.


\section{The nature of Albireo AB}

Finally, we can compare the systemic proper motion of Albireo \fat{Aa,Ac} (from the considerations in the previous section) with measured proper motions of Albireo B. This is done in the upper part of Fig.\,\ref{central}. The two violet circles labelled "B" denote the absolute proper motion of Albireo~B, as measured by Hipparcos and by Gaia DR2, respectively. They agree within their uncertainties. The green diamond represents the systemic motion of 
(+7.0/+4.6)~mas/year as derived above. Its uncertainty $\sigma$ is less than 2\,mas/year if the arguments of the previous section are valid. The difference between this tentative systemic motion and that of Albireo B is about 10\,mas/year, which is incompatible with zero at a level of at least 5$\sigma$. This motion difference corresponds to a difference in tangential velocity of 5\,($\pm$1)\,km/s at the Gaia DR2 parallax of 9.9\,($\pm$0.6)\,mas. All things taken into account, Albireo AB is quite probably  indeed an optical double, although not mainly for the reason given in the introduction.

Taking the difference in proper motion and in parallax at face value, and accepting that AB is not a physical pair, this still does not completely exclude a common genealogy: with just 5\,km/s in tangential velocity and 20\,pc in three-dimensional (3D)~space, their small separation in the combined five-dimensional (5D)~space is very unlikely for two unrelated galactic disk stars. Within the sphere of 100\,pc radius around the sun there are only a few B~stars, and their velocity scatter is of the order of 20\,km/s. Therefore, the probability of finding a pair of them within this volume and being so close in 5D~space is less than 10$^{-3}$. In other words, Albireo A and B might still be members of some dissolved star cluster or dissolving association.


\section{Discussion and proposed follow-up activities}
\label{discussion}

The above results on the mass of Albireo Aa and on the systemic proper motion of Albireo \fat{Aa,Ac} are obviously uncertain. This is mostly due to the uncertainty of the interferometric orbit of \fat{Aa,Ac}. Although the formal uncertainties of the orbital parameters by \citet{Scardia2008} are quite small, it is strange that the addition of just a few more observations at somewhat later epochs produced a completely different solution \citep{Roberts2018}. This leads to the suspicion that the \citet{Scardia2008}~solution might be nearly degenerate, and thus more uncertain than the formal errors indicate. \fat{The suspected near degeneracy is nicely illustrated by the similarity of the two orbit solutions over the time range covered by observations in Figs.\,\ref{Scardia-orbit} and\,\ref{Roberts-orbit}}.

However, a combination of various follow-up observations and other related activities would quickly remove the remaining uncertainties, and thus lead to final conclusions on both the astrophysics of the giant star Aa (i.e.~mass, evolutionary state, etc.) and on the nature of the wide double AB. It would be very useful to have as many of them as possible in about two years from now, when Gaia DR3 will provide even more precise astrometry, including instantaneous acceleration terms at about J2016 (see below). The following is a list of lines of investigation that would greatly advance our understanding of Albireo \fat{Aa,Ac}.  

\begin{itemize}
\item A few additional interferometric measurements at present epoch, even with moderate mas-level precision, would remove any doubt about the relative orbit of \fat{Aa,Ac}.
\item Radial-velocity measurements on the composite spectra of \fat{Aa,Ac}\footnote{We note that this cannot be done by the Gaia spectrograph, as the present 2D~angular separation of \fat{Aa,Ac} is not sufficiently well known. For the same reason, fibre-coupled spectrographs must be used in ground-based spectroscopy to avoid separation-induced velocity bias.} are complicated, but the enormous apparent brightness of the pair makes it very easy to get high-signal-to-noise-ratio(S/N) data even with small instruments. In combination with a consolidated interferometric orbital solution, a small series of precise radial-velocity measurements over the time interval till the publication of Gaia DR3 would provide independent mass estimates, for both components separately.
\item High-quality spectral classification, that is,~the determination of T$_{\rm eff}$, log\,g, and chemistry, is likewise complicated in composite spectra, but is not unfeasible. Again the brightness is helpful.
\item Precise values for T$_{\rm eff}$ along with the apparent magnitudes and the precisely known parallax would provide precise radii of both stars. These, in combination with log~g, would give a third independent determination of the masses. 
\item An extended astrometric solution of the raw Gaia observations of Albireo A should be performed, by adding acceleration terms (i.e.~time derivatives of the proper motion components in right ascension and declination). This would eliminate the disturbance of the DR2 parallax and motion by the currently existing astrometric acceleration, and would confirm or refute the long-term acceleration displayed by the red symbols in Fig.~\ref{central}. Along with an improved interferometric orbit, this would also provide a fourth independent piece of information on the mass of Aa.  
\item Such an extended astrometric Gaia solution could already be produced --- but only by the Gaia consortium --- using the existing (yet unpublished) individual astrometric observations which entered Gaia DR2. 
\item If the suggestion in the previous item cannot be followed, absolute acceleration terms will become available in 2021: Gaia DR3, to be released in just over two years, is announced to provide some binary-star solutions to the astronomical public. Special care should be taken by the Gaia consortium to ensure that such a solution for Albireo A  be included in DR3. At several tenths of a mas/(year)$^2$, the acceleration terms will be highly significant, that is,~they will have a very high relative precision.     
\item Finally, radial-velocity data on \fat{Aa,Ac}, along with the interferometric orbit, will give a consolidated systemic radial velocity, that is,~the sixth dimension to judge a possible genealogic connection between A and B. If the systemic radial velocity of \fat{Aa,Ac} is also close to that of B, a common origin would become very likely. In this case a search for more members of the dispersing cluster or association would become worthwhile. This can be done with Gaia DR2, or after the release of Gaia DR3 (then with many more available radial velocities for candidate members).
\end{itemize}

\fat{In the context of radial velocities it is interesting to note that  the radial velocities of Aa and B were equal in the General Catalogue of Radial Velocities (GCRV) of 1953 --- within the uncertainties set mainly by the broad spectral lines of B. This, however, could be quite different at a different orbital phase of the pair Aa,Ac. More recent values deviate from each other: -24.07\,($\pm0.12$)\,km/s for Aa \citep{Famaey2005}, and -18.80\,($\pm2.2$)\,km/s for B \citep{Kharchenko2007}. But this in turn could be entirely due to the orbital motion of Aa,Ac. New radial-velocity measurements are therefore clearly desirable.  }

\section{Exotic astrophysics?}

Can the existing photometric data and spectral type be reconciled with the \citet{Scardia2008} orbit, that is,~with the low mass estimate for Albireo Aa? This could be possible, if the K~star were a bloated low-mass star of 0.5\,$M_\odot$ for example, that had swallowed a 0.1\,$M_\odot$ companion 10\,000 years or so ago.

Can the existing photometric data and spectral type be reconciled with the \citet{Roberts2018} orbit, that is,~with the very high mass sum for Albireo A? This is also possible, if a massive black hole were added as an invisible component to the giant Aa. In this case the observed absolute angular acceleration of component~A would be only a small part of the much larger relative acceleration predicted by the \citet{Roberts2018} orbit solution.

However, in view of the obvious and relatively non-time-consuming possibilities to clarify the nature of the system observationally (see Section\,\ref{discussion} above), a deeper quantitative evaluation of such exotic scenarios does not seem useful. 

\begin{acknowledgements}
      This research has made use of the SIMBAD database, operated by the CDS at Strasbourg, France. \\
      This work has made use of data from the European Space Agency (ESA) mission {\it Gaia} (\url{https://www.cosmos.esa.int/gaia}), processed by the {\it Gaia} Data Processing and Analysis Consortium (DPAC, \url{https://www.cosmos.esa.int/web/gaia/dpac/consortium}). Funding for the DPAC has been provided by national institutions, in particular the institutions participating in the {\it Gaia} Multilateral Agreement.\\
      We thank U. Reichert, S. Jordan and K.P. Schr\"oder for valuable comments on a draft version of this paper.\\
      \fat{Furthermore, we thank our referee Brian Mason for valuable suggestions for improvement, especially of Figs.\,1 and\,2.} 
\end{acknowledgements}


\bibliographystyle{aa} 
\bibliography{Albireo} 

\end{document}